\documentclass[espf]{aa}
\usepackage{graphicx}

\usepackage{color} 

\usepackage{graphicx}
\usepackage{amsmath}
\usepackage[psamsfonts]{euscript}
\usepackage[psamsfonts]{amssymb}
\usepackage{amsmath}
\usepackage{graphicx}

\usepackage{epsfig}

\titlerunning{}

\def\kms{km\,s$^{-1}$}

\def\Mo{M_{\odot}}
\def\Lo{L_{\odot}}
\def\Ro{R_{\odot}}

\begin{document}

\title{Understanding the dynamical structure of pulsating stars: \\ The center-of-mass velocity and the Baade-Wesselink projection factor of the $\beta$~Cephei star $\alpha$ Lup \thanks{Based on observations made with ESO telescopes at the Silla Paranal Observatory under programme IDs 085.C-0614(A)}}

\titlerunning{The k-term and the Baade-Wesselink p-factor \\ of the $\beta$~Cephei star $\alpha$ Lup}
\authorrunning{Nardetto et al. }

\author{N. Nardetto \inst{1}, P. Mathias \inst{2}, A. Fokin \inst{1,3,4},  E. Chapellier \inst{1}, G. Pietrzynski \inst{4,5}, W. Gieren \inst{4},  \\ D. Graczyk \inst{4,5}, D. Mourard\inst{1}}

\institute{Laboratoire Lagrange, UMR7293, Universit\'e de Nice Sophia-Antipolis, CNRS, Observatoire de la C\^ote d'Azur, Nice, France Nicolas.Nardetto@oca.eu   \and Institut de Recherche en Astrophysique et Plan\'etologie, CNRS, 14 avenue Edouard Belin, Universit\'e de Toulouse, UPS-OMP, IRAP, 31400 Toulouse, France  \and Institute of Astronomy of the
Russian Academy of Sciences, 48 Pjatnitskaya Str., Moscow 109017 Russia  \and Universidad de Concepc\'ion, Departamento de Astronom\'ia, Casilla 160-C, Concepc\'ion, Chile  \and Warsaw University Observatory, Al. Ujazdowskie 4, 00-478, Warsaw, Poland}

\date{Received ... ; accepted ...}

\abstract{High-resolution spectroscopy of pulsating stars is a powerful tool to study the dynamical structure of their atmosphere. Lines asymmetry is used to derive the center-of-mass velocity of the star, while a direct measurement of the atmospheric velocity gradient helps to determine the projection factor used in the Baade-Wesselink methods of distance determination.} {We aim at deriving the center-of-mass velocity and the projection factor of the $\beta$ Cephei star $\alpha$ Lup.} {We present HARPS (High Accuracy Radial velocity Planetary Search) high spectral resolution observations of $\alpha$ Lup. We calculate the first-moment radial velocities and fit the spectral line profiles by a bi-Gaussian to derive line asymmetries. Correlations between the $\gamma$-velocity and the $\gamma$-asymmetry (defined as the average values of the radial velocity and line asymmetry curves respectively) are used to derive the center-of-mass velocity of the star. By combining our spectroscopic determination of the atmospheric velocity gradient with a hydrodynamical model of the photosphere of the star, we derive a semi-theoretical projection factor for $\alpha$~Lup.} {We find a center-of-mass velocity of $V_{\mathrm{\gamma}}=7.9 \pm 0.6$~\kms and that the velocity gradient in the atmosphere of $\alpha$~Lup is null. We apply to $\alpha$ Lup the usual  decomposition of the projection factor into three parts, $ p = p_\mathrm{0} f_\mathrm{grad} f_\mathrm{og} $ (originally developed for Cepheids), and derive a projection factor of $p=1.43\pm0.01$. By comparing our results with previous HARPS observations of classical Cepheids, we also point out a linear relation between the atmospheric velocity gradient and the amplitude of the radial velocity curve. Moreover, we observe a phase shift (Van Hoof effect), whereas $\alpha$~Lup has no velocity gradient. New HARPS data of a short-period $\beta$ Cephei star, $\tau^{1}$~Lup, are also presented in this paper.} {By comparing Cepheids and $\beta$ Cephei stars, these results bring insight into the dynamical structure of pulsating star atmospheres, which helps to better understand the k-term problem and the Baade-Wesselink p-factor for Cepheids. }


\maketitle

\section{Introduction}\label{s_Introduction}

A member of the $\beta$ Cephei variable class, $\alpha$ Lupi (HD 129056 = HR 5469, B1.5 III, V=2.3) is characterized by small light variations ($20 <  \Delta V < 30$ mmag; Van Hoof 1964, 1965), a radial velocity period of $P=0.2598$d or $3.85$ d$^{-1}$ (Heynderickx 1992), and a 2K-amplitude of about 20~\kms (Pagel 1956, Rodgers \& Bell 1962, Milone 1962, Van Hoof 1964). Lampens \& Goossens (1982) found that both the light variability and the radial velocity variations can be represented by an oscillation with a constant frequency. Then, performing a detailed spectroscopic analysis, Mathias et al. (1994) found, in addition to the main radial mode, a small-amplitude non-axisymmetric mode with degree $ 1 \leq l \leq 3$. This additional non-radial mode is considered as negligible in first approximation, which makes the comparison with classical Cepheids relevant and extremely interesting for various reasons.

First, the motion of Classical Cepheids in the Milky~Way is puzzling and has led to disagreements in the literature for decades. If an axisymmetric rotation of the Galaxy is taken into account, Cepheids appear to `fall' towards the Sun with a mean blue-shifted velocity of about $2$~\kms. This residual velocity shift has been dubbed the k-term and was first estimated by Joy (1939). A debate has raged for decades as to whether this phenomenon was truly related to the actual motion of the Cepheids and, consequently, to a complicated rotating pattern of our Galaxy, or if it was the result of effects within the atmospheres of the Cepheids (Camm 1938, 1944; Parenago 1945; Stibbs 1956; Wielen 1974; Caldwell \& Coulson 1987; Moffett \& Barnes 1987; Wilson et al.\ 1991; Pont, Mayor \& Burki 1994; Butler et al. 1996). In Nardetto et al.~(2008), we derived calibrated center-of-mass velocities of eight Cepheids observed with HARPS (High Accuracy Radial velocity Planetary Search) spectrograph using spectral line asymmetry.  By comparing these systemic velocities with the ones found in the literature (generally based on the cross-correlation method) and in particular in the Galactic Cepheid data base (Fernie et al. 1995), we obtained an average red-shifted correction of $1.8 \pm 0.2$~\kms. This result shows that the k-term of Cepheids stems from an intrinsic property of Cepheids. However, this physical explanation should be generalized to be reinforced. Studying the k-term for other kinds of pulsating stars (if it exists) is extremely interesting to identify its physical origin.  

Second, by comparing the dynamical structure of pulsating stars' atmospheres, one can seek interesting relations in the Hertzsprung-Russell (HR) diagram and produce constraints on the physical nature of the pulsation. The $\beta$ Cephei star $\alpha$ Lup is studied in this paper, while $\delta$~Scuti stars are presented in Guiglion et al. (2013).

Third, the Baade-Wesselink (hereafter \emph{BW}) method of determining the distances of Cepheids was recently used to calibrate the period-luminosity (\emph{PL}) of Galactic and Large Magellanic Cloud Cepheids (Fouqu\'e et al. 2007, Storm et al. 2011ab). The basic principle of this method is to compare the linear and angular size variation of a pulsating star in order to derive its distance through a simple division. The angular diameter is either derived by interferometry (e.g. Kervella et al. 2004; Davis et al. 2009) or by using the infrared surface brightness (hereafter IRSB) relation (Gieren et al. 1998, 2005a). However, when determining the linear radius variation of the Cepheid by spectroscopy, one has to use a conversion projection factor from radial to pulsation velocity.
This quantity has been studied using hydrodynamic calculations by Sabbey et al. (1995) and more recently by Nardetto et al. (2004, 2007, 2009, 2011). Conversely, the period-luminosity relation of $\beta$~Cephei stars (calibrated through parallax measurements) contains a significant scatter (McNamara \& Mathews 1967; Balona \& Feast 1975; Waelkens 1981; Sterken \& Jerzykiewicz 1992). The scatter is probably due to the finite width of the instability strip (Leung 1967; Jakate 1980; Bin Tian et al. 2003) and the presence of different modes among $\beta$~Cephei stars (Lesh \& Aizenman 1974). Although Sterken \& Jerzykiewicz (1979) have concluded on the basis of this scatter that a \emph{PL} relation does not exist for $\beta$~Cephei stars, deriving the projection factor is useful for different reasons. First, the projection factor can be used to derive the actual radius of the star and the acceleration in the atmosphere (Mathias et al.  1991, 1994). Moreover, the BW method can be applied to any kind of pulsating star in principle, even if the star pulsates in a non-radial mode (Dziembowski 1977, Balona \& Stobie 1979, Stamford \& Watson 1981, Hatzes 1996). However, pulsating star with shockwaves should be avoided (Mathias et al. 2006). Deriving the \emph{BW} distance of $\beta$ Cephei stars could help in calibrating the \emph{PL} relation and in better understanding the physical reason for the scatter. This paper is part of the international ``Araucaria~Project'', whose purpose is to provide an improved local calibration of the extragalactic distance scale out to distances of a few megaparsecs (Gieren et al. 2005b).

In Sect. 2, we present our HARPS observations of $\alpha$~Lup, which is the main interest of this paper, in addition to those of $\tau^{1}$~Lup. $\tau^{1}$~Lup  (HD 126341 = HR 5395, B2 IV, V=4.5) is a $\beta$~Cephei star of shorter period ($P=0.17736934$d or $5.637953$ d$^{-1}$;  Cuypers 1987), which we use in the analysis of Sect.~4 only. In Sect.~3, we derive the center-of-mass velocity of $\alpha$ Lup using spectral line asymmetry and compare it with previous estimates. Section~4 is devoted to a comparison of the dynamical structures of the atmospheres of $\beta$ Cephei stars ($\alpha$~Lup and $\tau^{1}$~Lup) and classical Cepheids. The results in Sects. 3 and 4 are based on observations only. In Sect. 5, we apply the projection factor decomposition to $\alpha$~Lup. We end with some conclusions (Sect.~6).

\section{Observations and analysis}

\begin{figure}[htbp]
\begin{center}
\resizebox{1.05\hsize}{!}{\includegraphics[clip=true]{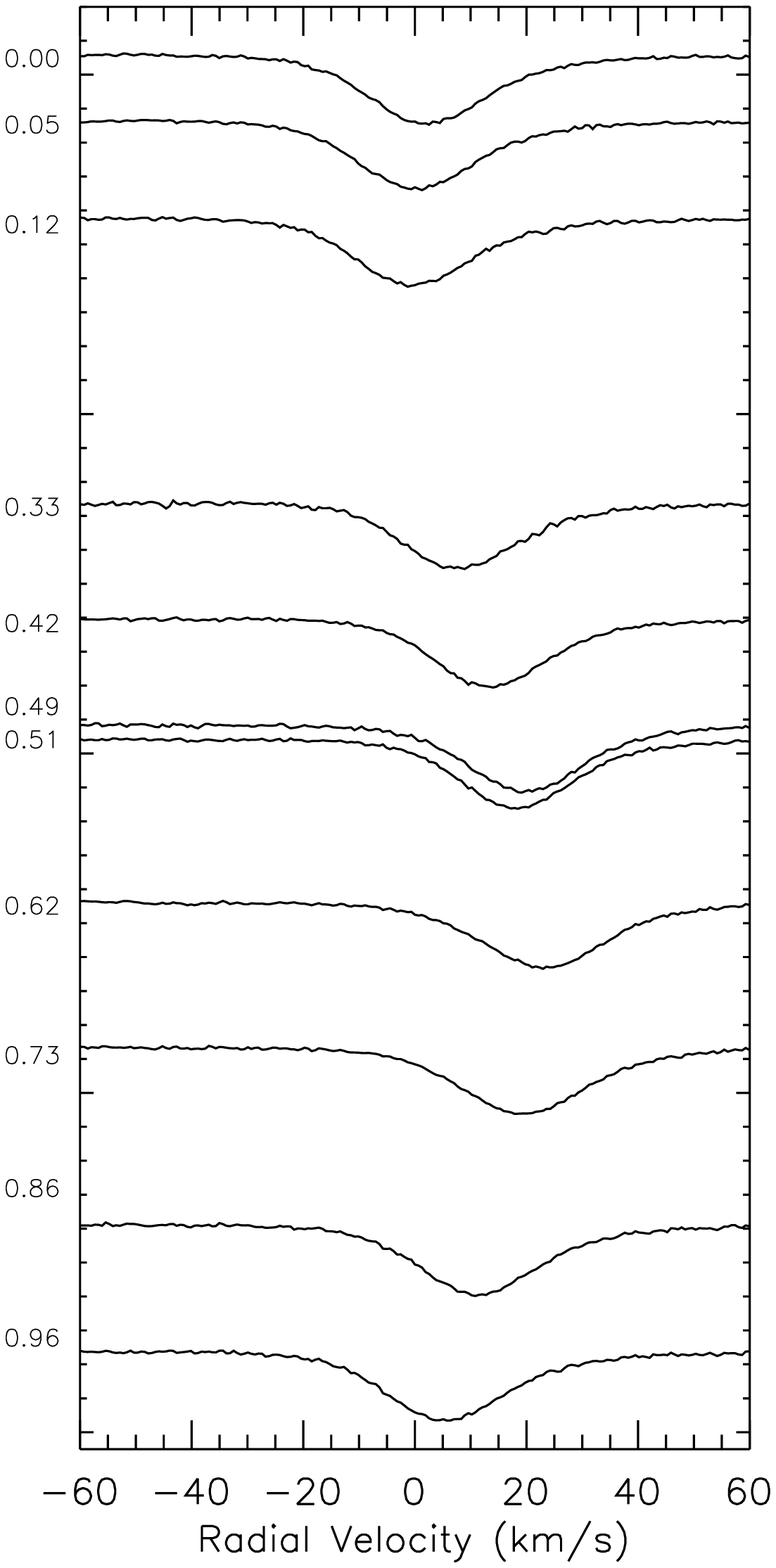}}
\end{center}
\caption{\ion{O}{II} 4705.3~\AA\  ($D \simeq 17$\%) spectral line evolution for $\alpha$~Lup. Pulsation
phases are given on the left of each profile. Wavelengths have been
translated into velocities (positive velocities
correspond to a redshift or receding motion).} \label{Figuv}
\end{figure}

\begin{figure}[htbp]
\begin{center}
\resizebox{1.05\hsize}{!}{\includegraphics[clip=true]{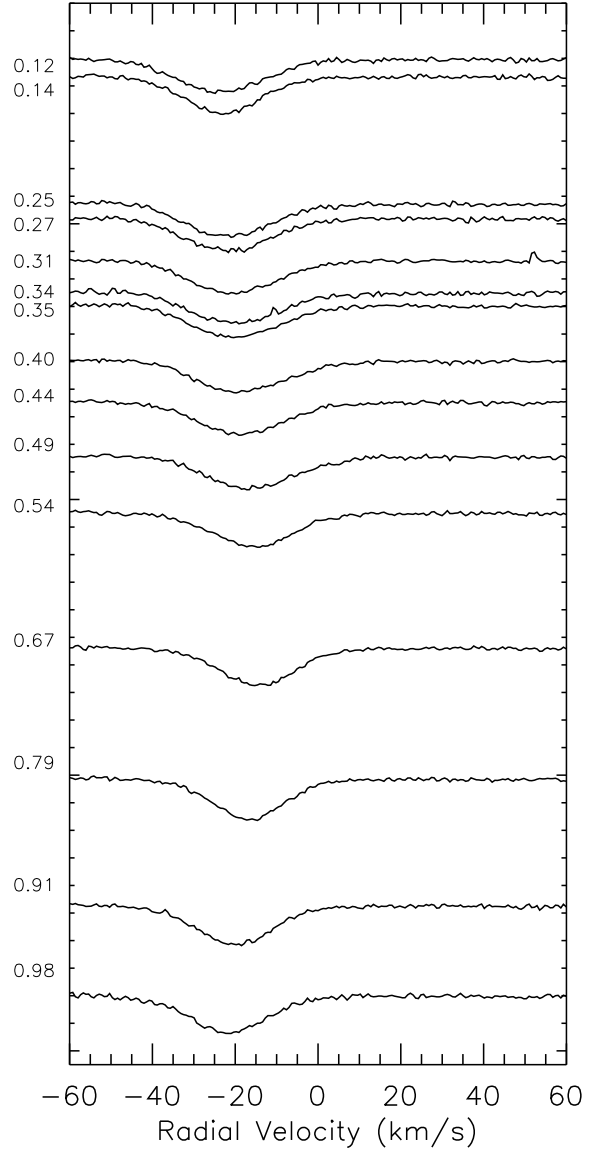}}
\end{center}
\caption{\ion{O}{II} 4705.3~\AA\  ($D \simeq 11$\%) spectral line evolution for $\tau^{1}$~Lup (same legend as in Fig. 1).} \label{Figuv}
\end{figure}

\begin{table*}
\begin{center}
\caption[]{HARPS observations results for $\alpha$ Lup and $\tau^1$ Lup for the \ion{O}{II} 4705.3~\AA\ spectral line (see also Figures 1 and 2).
\label{Tab_alpha_results}}
\begin{tabular}{ccccccccc}
\hline \hline \noalign{\smallskip}

  JD$_{\mathrm c}$  &  phase   &   Cy.     &      $ RV_{\mathrm c}$       &   $FWHM$   & $D$  & $A$ & $S/N$ & $\chi_{\mathrm red}^{2}$  \\
$[$d$]$                  &                &              & $[$\kms$]$                                 &   $[$\AA$]$                 &          &  $[$\%$]$     &             &          \\
  {\tiny (a)}  &  {\tiny (b)}   &   {\tiny (c)}      &   {\tiny (d)}       &   {\tiny (e)}   & {\tiny (f)}  & {\tiny (g)} & {\tiny (h)} & {\tiny (i)}  \\
\hline
\noalign{\smallskip}
\multicolumn{9}{c}{$\alpha$ Lup}\\
\noalign{\smallskip}
430.055	&	0.00	&	1	&	3.4	$\pm$	0.6	&	0.396	$\pm$	0.002	&	0.18	&	-10.00	$\pm$	0.81	&	356	&	2.2	\\
430.069	&	0.05	&	1	&	1.3	$\pm$	0.4	&	0.389	$\pm$	0.002	&	0.18	&	-9.33  	$\pm$	0.76	&	395	&	2.7	\\
430.087	&	0.12	&	1	&	1.0	$\pm$	0.5	&	0.385	$\pm$	0.002	&	0.17	&	-12.26	$\pm$	0.92	&	347	&	2.6	\\
431.961	&	0.33	&	8	&	9.4	$\pm$	0.7	&	0.377	$\pm$	0.003	&	0.17	&	-13.46	$\pm$	1.34	&	255	&	1.4	\\
431.984	&	0.42	&	8	&	14.4	$\pm$	0.5	&	0.367	$\pm$	0.002	&	0.18	&	-10.01	$\pm$	0.82	&	381	&	1.9	\\
430.963	&	0.49	&	4	&	20.4	$\pm$	0.8	&	0.376	$\pm$	0.002	&	0.18	&	-4.39 	$\pm$	1.00	&	298	&	1.6	\\
432.006	&	0.51	&	8	&	19.1	$\pm$	0.6	&	0.367	$\pm$	0.002	&	0.18	&	-7.65 	$\pm$	0.80	&	373	&	1.6	\\
430.997	&	0.62	&	4	&	23.3	$\pm$	1.0	&	0.404	$\pm$	0.002	&	0.17	&	1.54	         $\pm$	1.04	&	279	&	1.2	\\
431.025	&	0.73	&	4	&	20.1	$\pm$	0.8	&	0.400	$\pm$	0.002	&	0.17	&	-5.10 	$\pm$	0.94	&	315	&	1.0	\\
431.060	&	0.86	&	4	&	12.2	$\pm$	0.6	&	0.383	$\pm$	0.002	&	0.18	&	-6.50	         $\pm$	0.80	&	359	&	2.5	\\
431.084	&	0.96	&	4	&	6.7	$\pm$	0.5	&	0.389	$\pm$	0.002	&	0.18	&	-10.06	$\pm$	0.79	&	387	&	2.3	\\
\noalign{\smallskip}
\multicolumn{9}{c}{$\tau^1$ Lup}\\
\noalign{\smallskip}
431.091	&	0.12	&	7	&	-23.32	$\pm$	1.25	&	0.388	$\pm$	0.004	&	0.11	&	-2.49		$\pm$	2.42	&	200	&	0.8	\\
431.981	&	0.14	&	12	&	-22.77	$\pm$	1.18	&	0.337	$\pm$	0.004	&	0.12	&	0.48		$\pm$	2.41	&	191	&	0.8	\\
430.049	&	0.25	&	1	&	-20.48	$\pm$	0.90	&	0.381	$\pm$	0.005	&	0.11	&	-9.50		$\pm$	2.23	&	225	&	1.3	\\
432.003	&	0.27	&	12	&	-21.07	$\pm$	1.27	&	0.384	$\pm$	0.005	&	0.11	&	-1.91		$\pm$	2.58	&	187	&	0.9	\\
430.060	&	0.31	&	1	&	-19.78	$\pm$	1.09	&	0.378	$\pm$	0.006	&	0.11	&	-7.96		$\pm$	2.80	&	186	&	0.5	\\
432.017	&	0.34	&	12	&	-19.06	$\pm$	1.01	&	0.388	$\pm$	0.005	&	0.10	&	-7.25		$\pm$	2.53	&	214	&	1.4	\\
430.066	&	0.35	&	1	&	-19.25	$\pm$	0.76	&	0.384	$\pm$	0.004	&	0.11	&	-8.73		$\pm$	1.98	&	269	&	0.9	\\
430.075	&	0.40	&	1	&	-17.61	$\pm$	0.81	&	0.392	$\pm$	0.006	&	0.11	&	-16.34	$\pm$	2.31	&	256	&	0.8	\\
430.969	&	0.44	&	6	&	-18.00	$\pm$	1.14	&	0.385	$\pm$	0.004	&	0.11	&	-4.22		$\pm$	2.26	&	216	&	0.8	\\
430.091	&	0.49	&	1	&	-15.44	$\pm$	0.78	&	0.391	$\pm$	0.005	&	0.11	&	-13.26	$\pm$	2.24	&	247	&	1.4	\\
430.987	&	0.54	&	7	&	-15.64	$\pm$	0.83	&	0.359	$\pm$	0.004	&	0.11	&	6.46		$\pm$	2.09	&	233	&	0.7	\\
431.011	&	0.67	&	7	&	-14.57	$\pm$	0.79	&	0.333	$\pm$	0.003	&	0.12	&	3.38		$\pm$	1.92	&	235	&	0.9	\\
431.031	&	0.79	&	7	&	-16.81	$\pm$	0.81	&	0.309	$\pm$	0.003	&	0.13	&	2.79		$\pm$	1.77	&	244	&	1.0	\\
431.053	&	0.91	&	7	&	-19.85	$\pm$	1.24	&	0.340	$\pm$	0.004	&	0.13	&	3.61		$\pm$	2.15	&	196	&	0.8	\\
431.068	&	0.99	&	7	&	-21.73	$\pm$	1.49	&	0.351	$\pm$	0.004	&	0.13	&	1.83		$\pm$	2.59	&	168	&	0.9	\\
 \hline \hline \noalign{\smallskip}
\end{tabular}
\end{center}

\begin{list}{}{}
\item {\tiny (a)} JD$_{\mathrm c}$, average Julian date of
observation defined by $JD_{\mathrm c}=JD-2455000$ [in days].

\item {\tiny (b)}  pulsation phase of
observation using the ephemeris indicated in the text.

\item {\tiny (c)} Cy., pulsating cycle of the star
corresponding to observation.

\item {\tiny (d)} $ RV_{\mathrm c}$, heliocentric radial velocity
corresponding to the first moment of the spectral line [in
\kms].

\item {\tiny (e)} $FWHM$, full-width at half-maximum
derived from the bi-Gaussian fit [in Angstroms].

\item {\tiny (f)} $D$, line depth derived from the
bi-Gaussian fit [no dimension]. Errors bars are not indicated but are of
the order of $10^{-4}$.

\item {\tiny (g)} $A$, asymmetry derived from the
bi-Gaussian fit [in percentage].

\item {\tiny (h)} $S/N$, signal-to-noise ratio calculated in the continuum next to the  \ion{O}{II} 4705.3~\AA\  spectral line.

\item {\tiny (i)} $\chi_{\mathrm red}^{2}$, reduced  $\chi^{2}$ factor corresponding to the bi-Gaussian fit.

\end{list}
\end{table*}

\begin{table}
\begin{center}
\caption{Spectral lines used in this study with corresponding $\gamma$-velocity ($V_\gamma$) and  $\gamma$-asymmetry ($A_\gamma$) obtained for $\alpha$ Lup. \label{Tab_Lines}}
\begin{tabular}{llcc}
\hline \hline \noalign{\smallskip}
 Elements & $\lambda$ (\AA) & $V_\gamma$ (\kms)& $A_\gamma$ (\%) \\
\hline
\ion{Si}{II} &      3856.02 &$        11.81 \pm       1.36 $&$       -7.41 \pm        3.52 $\\
\ion{O}{II} &       3907.44 &$        10.78 \pm       0.96 $&$       -6.52 \pm        6.92 $\\
\ion{O}{II} &       3911.96 &$        11.28 \pm       0.98 $&$       -11.92\pm        1.52 $\\
\ion{O}{II} &       3945.03 &$        10.73 \pm       0.95 $&$       -8.49 \pm        2.00 $\\
\ion{O}{II} &       3954.36 &$        9.90  \pm       1.33 $&$       -8.15 \pm        2.69 $\\
\ion{N}{II} &       3955.85 &$        9.50  \pm       0.85 $&$       -7.25 \pm        2.49 $\\
\ion{O}{II} &       3982.71 &$        9.79  \pm       0.70 $&$       -4.49 \pm        1.73 $\\
\ion{N}{II} &       3995.00 &$        9.22  \pm       1.44 $&$       -5.89 \pm        1.07 $\\
\ion{N}{II} &       4041.31 &$        9.59  \pm       1.11 $&$       -6.34 \pm        2.03 $\\
\ion{Si}{II} &       4128.05 &$        10.32 \pm       0.92 $&$       -8.21 \pm        2.18 $\\
\ion{Fe}{III} &       4137.76 &$        10.45 \pm       0.62 $&$       -11.33\pm        3.16 $\\
\ion{S}{II} &       4162.67 &$        8.96  \pm       0.87 $&$       -3.94 \pm        4.82 $\\
\ion{Fe}{III} &       4166.84 &$        9.88  \pm       0.55 $&$       -10.25\pm        4.94 $\\
\ion{N}{II} &       4171.60 &$        10.53 \pm       0.58 $&$       -13.50\pm        4.28 $\\
\ion{O}{III} &       4185.44 &$        10.42 \pm       1.03 $&$       -7.52 \pm        2.74 $\\
\ion{N}{II} &       4227.74 &$        9.17  \pm       0.63 $&$       -8.35 \pm        2.56 $\\
\ion{S}{III} &       4361.53 &$        9.71  \pm       0.60 $&$       -5.06 \pm        3.00 $\\
\ion{O}{II} &       4414.88 &$        10.94 \pm       0.96 $&$       -6.56 \pm        0.69 $\\
\ion{O}{II} &       4416.97 &$        10.17 \pm       1.43 $&$       -6.76 \pm        1.22 $\\
\ion{Fe}{III} &       4419.60 &$        9.56  \pm       0.80 $&$       -6.76 \pm        2.41 $\\
\ion{Ar}{II} &       4426.00 &$        9.06  \pm       0.42 $&$       -1.14 \pm        4.44 $\\
\ion{He}{II} &       4437.55 &$        11.17 \pm       1.04 $&$       -5.56 \pm       0.88 $\\
\ion{O}{II} &       4452.38 &$        9.36  \pm       0.58 $&$       -6.03 \pm        1.48 $\\
\ion{Al}{III} &       4512.56 &$        8.97  \pm       0.69 $&$       -4.77 \pm        1.70 $\\
\ion{Si}{III} &       4567.84 &$        10.61 \pm       0.72 $&$       -8.69 \pm       0.31 $\\
\ion{Si}{III} &       4574.76 &$        10.03 \pm       0.74 $&$       -8.15 \pm       0.51 $\\
\ion{O}{II} &      4590.97 &$        10.03 \pm       0.60 $&$       -8.11 \pm       0.72 $\\
\ion{N}{II} &       4607.15 &$        9.78  \pm       0.54 $&$       -6.04 \pm       0.93 $\\
\ion{O}{II} &       4641.81 &$        10.70 \pm       0.88 $&$       -8.00 \pm       0.69 $\\
\ion{O}{II} &       4661.63 &$        10.14 \pm       0.64 $&$       -6.75 \pm       0.67 $\\
\ion{O}{II} &       4676.23 &$        10.16 \pm       0.98 $&$       -7.34 \pm        1.31 $\\
\ion{O}{II} &       4705.32 &$        11.95 \pm       0.64 $&$       -8.17 \pm       0.91 $\\
\ion{Si}{III} &       4813.33 &$        10.68 \pm       0.51 $&$       -6.50 \pm        1.94 $\\
\ion{S}{II} &       4815.55 &$        9.52  \pm       0.38 $&$       -3.91 \pm        3.29 $\\
\ion{N}{II} &       5005.15 &$        10.19 \pm       0.84 $&$       -7.90 \pm       0.90 $\\
\ion{He}{I} &       5015.68 &$        9.49  \pm       0.57 $&$       -4.38 \pm       0.36 $\\
\ion{N}{II} &       5666.63 &$        9.12  \pm       0.67 $&$       -4.21 \pm       0.78 $\\
\ion{N}{II} &       5676.02 &$        9.42  \pm       0.67 $&$       -2.92 \pm        1.02 $\\
\ion{N}{II} &       5679.55 &$        10.06 \pm       0.83 $&$       -5.94 \pm       0.55 $\\
\ion{N}{II} &       5686.21 &$        9.92  \pm       0.72 $&$       -9.75 \pm        2.13 $\\
\ion{Al}{III} &       5696.60 &$        9.23  \pm       0.73 $&$       -3.91 \pm       0.63 $\\
\ion{N}{II} &       5710.77 &$        9.10  \pm       0.61 $&$       -5.74 \pm        1.42 $\\
\ion{Al}{III} &       5722.73 &$        9.22  \pm       0.83 $&$       -3.90 \pm       0.95 $\\
\ion{Si}{III} &       5739.73 &$        11.75 \pm       0.97 $&$       -10.07\pm       0.57 $\\
\ion{C}{II} &       6578.05 &$        9.31  \pm       0.89 $&$       -6.99 \pm       0.47 $\\
\ion{C}{II} &       6582.88 &$        8.95  \pm       0.80 $&$       -5.02 \pm       0.56 $\\
\hline \noalign{\smallskip}
\end{tabular}
\end{center}
\end{table}

The HARPS spectrometer is dedicated to the search for extrasolar
planets by means of radial velocity measurements. It is installed at
the Coud\'e room of the $3.6$ meter telescope at La Silla. The
resolution is $R=80000$ (in the EGGS mode) and the average signal to noise ratio we
obtain over all observations in the continuum is $340$
per pixel for $\alpha$~Lup  (11 spectra) and $220$ for $\tau^{1}$~Lup (15 spectra). We have used the standard ESO/HARPS pipeline reduction package. Using the Vienna Atomic Line Database\footnote{http://www.astro.uu.se/~vald/php/vald.php} (Piskunov et al. 1995 ; Heiter et al. 2008), we have identified 46 unblended spectral lines in the spectra of $\alpha$~Lup (Table~2).  Three lines were rejected in the analysis of $\tau^{1}$~Lup:  \ion{Si}{II} 3856.02~\AA, \ion{O}{II} 3907.44~\AA, and \ion{N}{II}  3995.00~\AA\ because of blends.

Following the same methods as described in Nardetto et al. (2006a), we derived for each spectral line of $\alpha$~Lup the first moment radial velocity, the full-width at half-maximum, the line depth, and the bi-Gaussian line asymmetry. The pulsation period given in the literature for  $\alpha$~Lup is $P=0.2598466$d (Heynderickx, 1992). We decided to take the Julian date of our first observation in our sample as the reference ($T_0=2455430.0553$d). Our 11 observations of $\alpha$~Lup spread over eight pulsating cycles. No significant differences (in radial velocity and in line asymmetry) were found from one cycle to the other, but more spectroscopic observations of this star would be necessary to confirm this statement. Nevertheless, using the FAMIAS software\footnote{Result obtained with the software package FAMIAS developed in the framework of the FP6 European Coordination Action HELAS (http://www.helas-eu.org/)} (Zima 2008) and the modes identified by Mathias et al. (1994), we found that the non-radial mode (compared to a {\it pure} radial mode) might lead to a systematic change of the amplitude of the radial velocities (and also line asymmetries) by only one percent (this is in principle true for all spectral lines). We consider in the following that the contamination of the data by a weak non-radial mode is negligible for our purposes. Therefore, spectra have been recomposed consistently into a unique cycle. 

For $\tau^1$~Lup we obtained 15 observations covering 12 cycles. We considered $P=0.17736934$d (Cuypers 1987) and $T_0=2455429.8273$d, such that the maximum of the radial velocity curve for $\alpha$~Lup and $\tau^{1}$~Lup corresponds to the same phase of pulsation ($\phi \simeq 0.3$). Again, no significant differences were found in the radial velocities from one cycle to the other, and the spectra have been recomposed consistently into a unique cycle to derive the radial velocity curve. However,  we found that the spectral line asymmetry is probably contaminated by additional modes which are eventually non-radial. While the latter are again negligible in the radial velocity determination (first moment of the spectral line profile), they are significative in the spectral line asymmetry (which can also be considered as the third moment of the spectral line profile). Consequently, we use in this study the radial velocity only and, the detailed analysis of the spectral line asymmetry and mode idenfication for $\tau^{1}$~Lup will be done in another paper (for this purpose, additional data should be obtained).


Figures 1 and 2 present the typical \ion{O}{II} 4705.3~\AA\  spectral line variation for $\alpha$~Lup and $\tau^{1}$~Lup respectively. Table 1 shows the results derived for both stars from the same line.

For $\alpha$ Lup, the $RV_{\mathrm{c}}$ and $A$ quantities for all selected spectral lines are interpolated over the pulsation phase using a periodic cubic spline function. Then, $A_{\gamma}$ and $V_{\gamma}$ are calculated by averaging the $A(\phi)$ and $RV_{\mathrm{c}}(\phi)$ interpolated curves, with $\phi$ the pulsation phase. The uncertainties on $A_{\gamma}$ and $V_{\gamma}$ are defined as the average values of individual uncertainties on $A$ and $RV_{\mathrm{c}}$ respectively. In Table 2 $A_{\gamma}$ and $V_{\gamma}$ are given in the case of $\alpha$ Lup. For $\tau^{1}$ Lup, we made a careful interpolation of  the $RV_{\mathrm{c}}$ quantities (used in Sect. 3). The analysis presented in next section, which is based on spectral line asymmetries, is applied to $\alpha$ Lup only.

\section{Center-of-mass velocity of $\alpha$ Lup}

\begin{figure*}[htbp]
\begin{center}
\resizebox{\hsize}{!}{\includegraphics[clip=true]{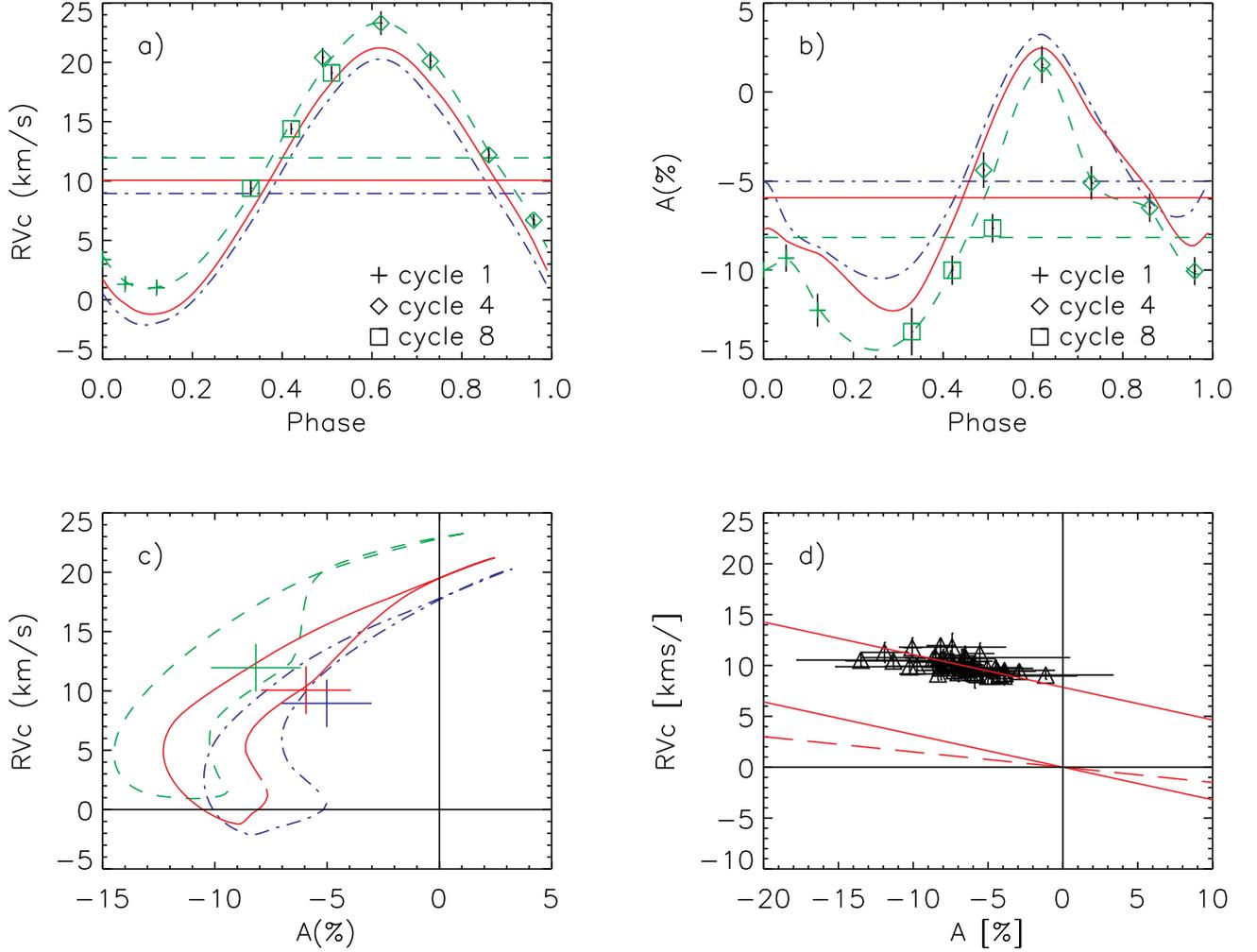}}
\end{center}
\caption{Illustration of the method applied to derive the center-of-mass velocity of $\alpha$ Lup from the line asymmetries (see text).} \label{Figuv}
\end{figure*}
\begin{figure}[htbp]
\begin{center}
\resizebox{0.96\hsize}{!}{\includegraphics[clip=true]{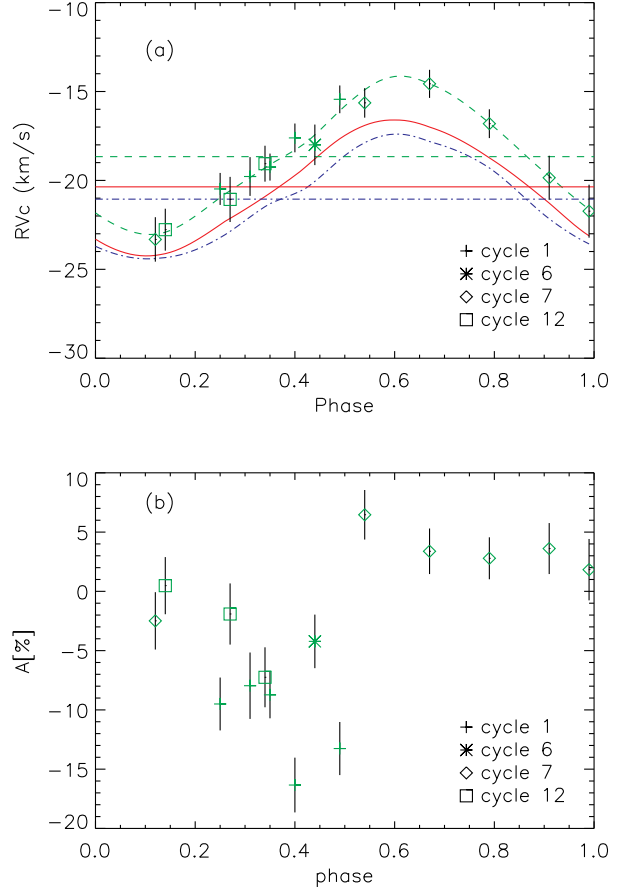}}
\end{center}
\caption{(a) Same as Fig. 3b in the case of $\tau^1$ Lup. (b) Spectral line asymmetry measurements for the \ion{O}{II} 4705.3~\AA\  spectral line. The cycles of pulsation are indicated (see Table 2). } \label{Figuv}
\end{figure}  

\begin{figure}[htbp]
\begin{center}
\resizebox{\hsize}{!}{\includegraphics[clip=true]{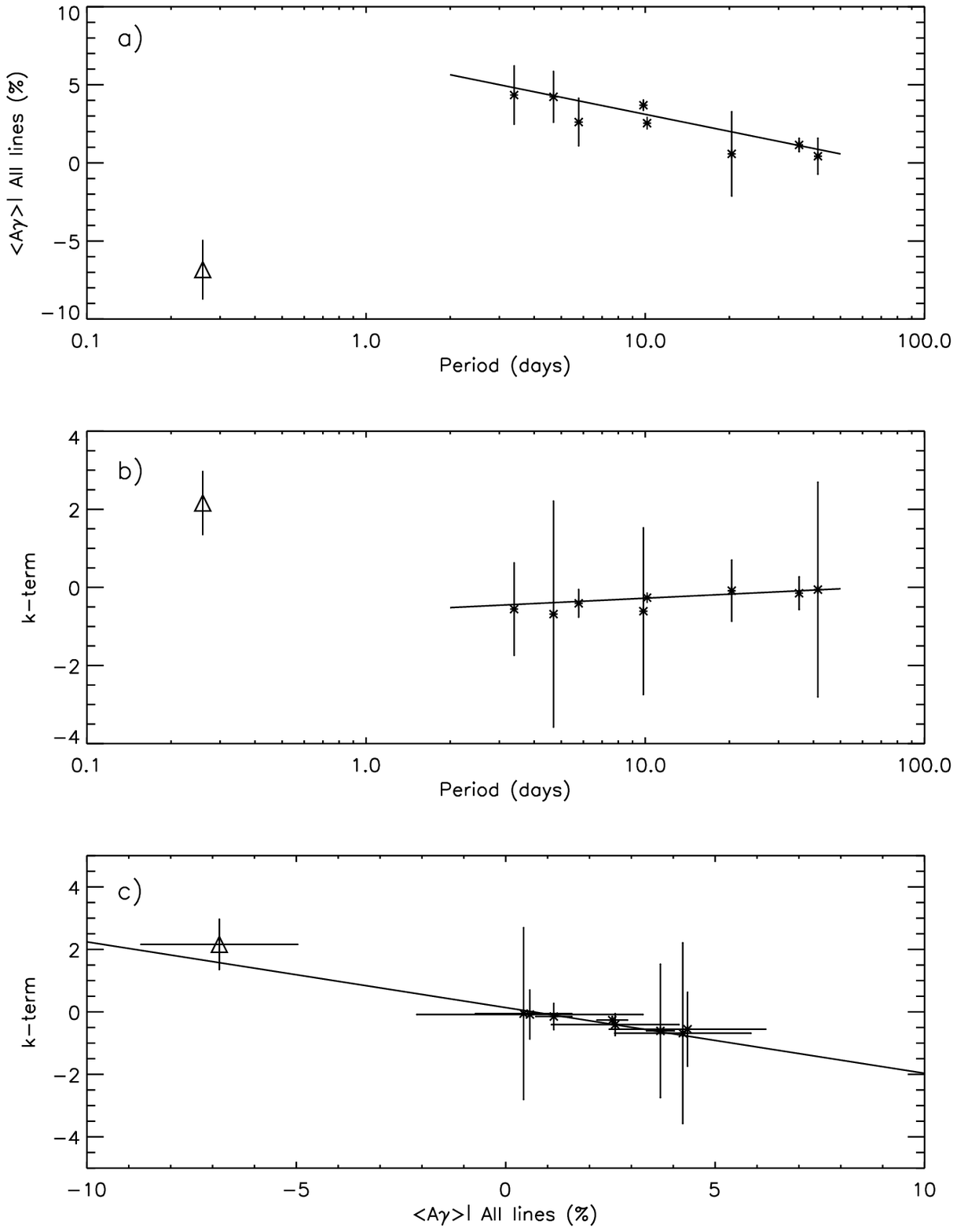}}
\end{center}
\caption{$\gamma$-asymmetry (averaged over all spectral lines) and the k-term (as defined in the text) as a function of the pulsation period of the star (in days).  The correlation between the k-term and the  $\gamma$-asymmetry is presented in panel c). The crosses are for the eight Cepheids studied in Nardetto et al. (2008) and the triangle is for $\alpha$ Lup. } \label{Figuv}
\end{figure}

The aim of this section is to study the $\gamma$-velocities ($V_{\gamma}$) defined as the average values of the radial velocity curve and to show that they consist of two components: one related to the space motion of the star itself and one (which we call the k-term) related to the dynamical structure of the star's atmosphere. For an introduction to our methodology (see also Nardetto et al. 2008), we refer to Fig. 3. 

Figures 3a and 3b show the interpolated radial velocity and asymmetry curves for three spectral lines:   \ion{O}{II}~4705.3~\AA\  (dashed line), \ion{N}{II}~5679.5~\AA\ 
(solid line), and \ion{C}{II}6582.9~\AA\  (dot-dashed line). The actual measurements (crosses etc.) are only indicated in the case of the \ion{O}{II} 4705.3~\AA\  line. From the interpolated curves, we now calculate the $\gamma$-velocities and $\gamma$-asymmetries corresponding to each spectral line (horizontal lines). An anti-correlation is clearly seen: the larger the $\gamma$-velocity, the lower the $\gamma$-asymmetry (for a given spectral line). 

As a comparison, we show in Fig. 4a the radial velocity curves obtained for $\tau^{1}$ Lup for the same spectral lines (same legend). Figure 4b shows the \ion{O}{II} 4705.3~\AA\  spectral line asymmetry as a function of the pulsation phase.  As said, we find that the spectral line asymmetry is probably contaminated by a multi-periodicity (or non-radial pulsations), while this is not the case for the radial velocity. 

In Fig.~3c, $RV_{\mathrm{c}}-A$ plots and the corresponding ($A_{\gamma}$,$V_{\gamma}$) average values (large crosses) for the three different lines are presented in the case of $\alpha$~Lup. Although the $RV_{\mathrm{c}}-A$ plots have different shapes (confirmed for all spectral lines), which is the result of the dynamical structure of the $\beta$ Cephei star atmosphere, the average values (large crosses) are aligned, confirming the anti-correlation found in Fig.~3ab. 

Fig.~3d is a generalization of diagram (c) for all spectral lines. The $RV_{\mathrm{c}}-A$ curves are not included for clarity. Upper values correspond to residual $\gamma$-velocities $V_{\gamma}(i)$ of the 46 selected spectral lines $i$. A linear fit is performed, and we find the relation

\begin{equation}
V_{\gamma}(i)=a_{\mathrm{0}} A_{\gamma}(i) + b_{\mathrm{0}},
\end{equation}
with $a_{\mathrm{0}}=-0.32 \pm 0.08$~\kms and $b_{\mathrm{0}}=7.86 \pm 0.59$~\kms. The origin of the plot is taken as a reference for the determination of the center-of-mass velocity of the star: the $\gamma$-velocity is assumed to be zero when the $\gamma$-asymmetry is zero. This means that all points

\begin{equation}
(A_{\gamma}(i), V_{\gamma}(i))
\end{equation}
are translated into
\begin{equation}
(A_{\gamma}(i), V_{\gamma}(i)-b_{\mathrm{0}}), 
\end{equation}
which allows the definition of a physically calibrated center-of-mass velocity for $\alpha$ Lup:
\begin{equation}
V_{\gamma\star}=b_{\mathrm{0}}=7.86\pm0.59 \:\mathrm{km.s}^{-1} \, .
\end{equation}

This quantity is relatively precise due to the very high S/N and spectral resolution of HARPS data. In Fig.~3d, the dashed line is the resulting $V_{\gamma}A_{\gamma}$ line ($a_{\mathrm{0}}=-0.15 \pm 0.01$) averaged over the eight Cepheids studied in Nardetto et al. (2008). The slopes obtained for Cepheids and $\alpha$ Lup are consistent at the 2$\sigma$ level. 

We now compare our center-of-mass velocity of $\alpha$ Lup to previous estimates, which yields some conclusions concerning the k-term phenomenon. First of all, we have to define several quantities. Originally, the k-term was defined as the difference between the expected systemic velocity of a Cepheid (due to the rotation of the Milky Way) and its measured $\gamma$-velocity (i.e. the average value of the interpolated radial velocity curve). Then, Nardetto et al.~(2008) estimated the k-term by comparing the measured $\gamma$-velocities found in the literature with the center-of-mass velocity of a few Cepheids derived with the method described in this section (based on spectral line asymmetry). They concluded that $\gamma$-velocities found in the literature (mainly from the Galactic Cepheid data set; Fernie et al. 1995) were probably biased by the effect of spectral line asymmetry and that a redshift correction of $2$~\kms was necessary.  We now apply the same analysis to $\alpha$ Lup. In Tab. 3, we compare our result $V_{\gamma\star}=7.9\pm0.6 \:\mathrm{km\: s}^{-1} \, $ with values in the literature. The dispersion of $\gamma$-velocities in the literature is too large (from 4.0 to 10.0\kms) to provide any conclusion (we do not find an offset of 2~\kms as was the case for Cepheids). This result is not surprising: The dispersion in the results presented in Tab.~3 comes mainly from the method used to derive the radial velocity, the method used to perform the interpolation, the considered spectral lines, or the data quality.

Finally, it seems that the most consistent way to study the k-term (and to proceed to the comparison of Cepheids with $\alpha$~Lup) is to understand the physical nature of the $\gamma$-asymmetry (or $\gamma$-velocity effects) by considering the following quantity (that we now call the k-term in the remaining paper): 

\begin{equation}
 k =  <V_{\gamma}> - V_{\gamma\star} = 10.1_{\pm 0.8} - 7.9_{\pm0.6} = 2.2_{\pm0.8} \:\mathrm{km\: s}^{-1} \, , 
\end{equation}
where $<V_{\gamma}>$ is defined as the average of the $V_{\gamma}$ quantities over all spectral lines in our sample (see Fig. 3d).  The value $<V_{\gamma}> = 10.1\pm0.8$~\kms should be comparable with the most recent observations (of good quality) obtained by Mathias et al. (1994), but this is not the case at the 3.5$\sigma$ level. As mentioned, it probably comes from the methods used. Similarly, we define $ <A_{\gamma}>  = -6.8\pm1.9 \%\, $ as the average of the $A_{\gamma}$ a quantities over all spectral lines in our sample.

Figure 5 shows this quantity $k$ and also the $\gamma$-asymmetry  $<A_{\gamma}>$ for $\alpha$ Lup and for the Cepheids analysed in Nardetto et al. (2008) as a function of the period of pulsation. Surprisingly, $\alpha$ Lup shows a mean negative $\gamma$-asymmetry of  $-6.8\pm1.9$\%, which is significantly lower compared to the positive and period-dependent values found for Cepheids (Fig. 5a). This spectral line asymmetry effect translates directly into a positive $\gamma$-velocity of $2.2\pm 0.8$\kms, while for Cepheids we found values between $-1$ and $0$ \kms (Fig. 5b). 

A first conclusion is that the k-term (as defined above) is correlated to the $\gamma$-asymmetry measurements (Fig. 5c),  confirming that it reflects an intrinsic property of pulsating stars. Moreover, these spectral line asymmetry effects seem to have different behaviour in $\alpha$ Lup and in classical Cepheids. More observations for stars covering a period between 0.2 and 2.0 days would be extremely interesting. Presently, there is no hydrodynamical model of pulsating stars able to reproduce these physical effects (i.e. $A_{\gamma}$) properly.

\begin{table}
\begin{center}
\caption{$\gamma$-velocities of $\alpha$ Lup derived in the literature \label{Tab_Lines}}
\begin{tabular}{llccc}
\hline \hline \noalign{\smallskip}
Reference & date & N measures & $V_\gamma$ & note\\
\hline
Wright &1909 & 10 &$7.8$ \\
Campbell &1928 & 16 & $7.0$ \\
Wilson &1953 & 16 &  $7.3$ \\
Pagel &1956& 70 &$ {\it 4.0 }\pm {\it 3.0}$ & {\it a}\\
Rogers &1962 & 24 & ${\it 10.0} \pm {\it 2.0}$ & {\it b}\\
Thackeray &1966 & 9 & $10.6 $ \\
Mathias &1994 & 450 &  ${\it 7.4} \pm {\it 1.7}$ & {\it c} \\
Barbier-Brossat &2000 & 133 & $5.4\pm 0.6$ & {\it d} \\

\hline
This work    &2012& 12 & $ 10.1 \pm 0.8$ \\
\multicolumn{2}{l}{ This work  : $V_\gamma \star$} & 11 & $7.9\pm0.6 $\\
\hline \noalign{\smallskip}
\end{tabular}
\end{center}

\begin{list}{}{}
\item {\tiny (a)} Estimated from their Fig. 6
\item {\tiny (b)} Estimated from their Fig. 1
\item {\tiny (c)} Estimated from their Tab. 1 (average over eight metallic lines). The precision for each individual line is $0.7$~\kms
\item {\tiny (d)} From the original data set of Evans (1967)

\end{list}

\end{table}

\section{The dynamical structure of the atmosphere of $\beta$ Cephei stars compared to classical Cepheids. }

\begin{figure}[htbp]
\begin{center}
\resizebox{1.05\hsize}{!}{\includegraphics[clip=true]{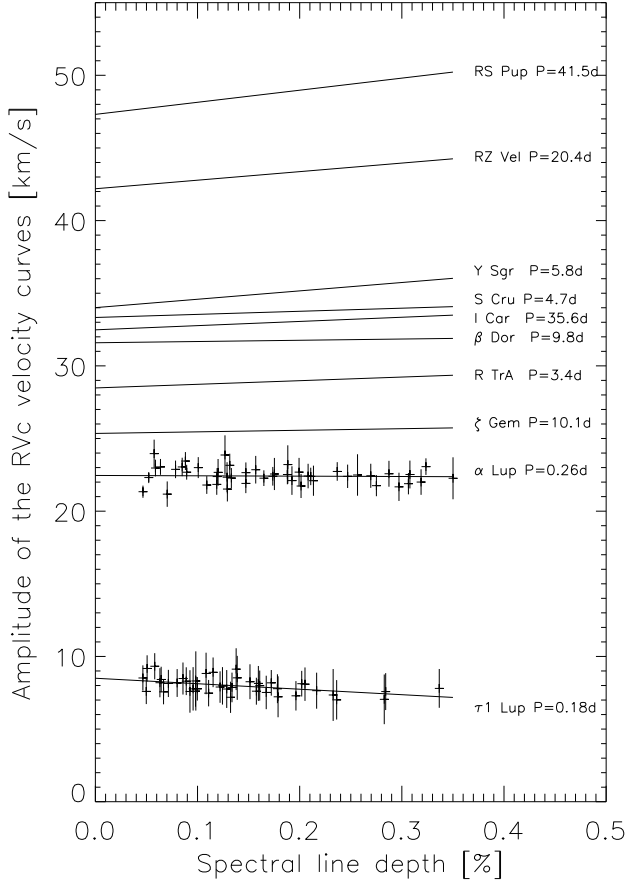}}
\end{center}
\caption{Amplitude of the $RV_{\mathrm{c}}$ curves as a function of the spectral line depth in the case of the $\beta$ Cephei stars $\alpha$ Lup and $\tau^{1}$ Lup (dots are the HARPS measurements presented in this paper with the corresponding uncertainties) and for the eight Cepheids studied in Nardetto et al. (2007).  } \label{Figuv}
\end{figure}

\begin{figure}[htbp]
\begin{center}
\resizebox{1.0\hsize}{!}{\includegraphics[clip=true]{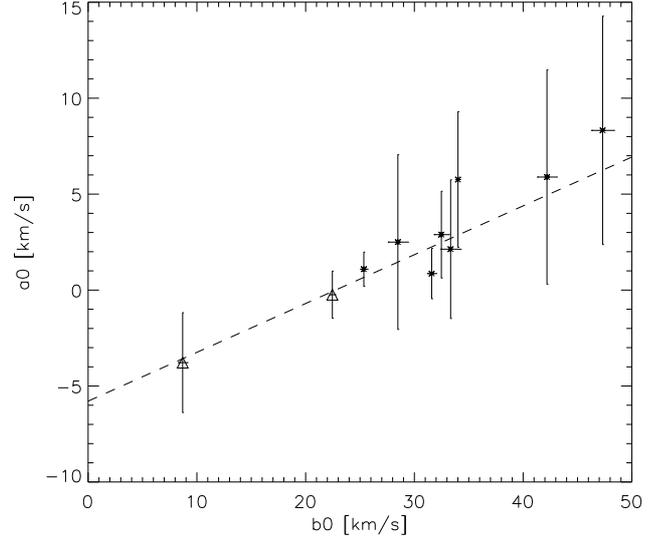}}
\end{center}
\caption{velocity gradient in the atmosphere of the stars as a function of the amplitude of the $RV_{\mathrm{c}}$ radial velocity curves. The crosses are for the eight Cepheids and the triangles are for $\beta$ Cephei stars, $\alpha$ Lup and $\tau^{1}$ Lup.} \label{Figuv}
\end{figure}

Using our HARPS data of $\beta$ Cephei stars ($\alpha$~Lup and $\tau^{1}$~Lup) and our previous data obtained for Cepheids, we try to point out some differences in terms of atmospheric velocity gradients and phase shifts. 

In Nardetto et al. (2007), we have shown that the line depth (taken at the minimum radius phase, hereafter $D$) is a good indicator of the line-forming regions. In this case, the photosphere corresponds to a zero line depth. By comparing the 2K amplitude (defined as the amplitude of the first moment radial velocity curve, hereafter $ \Delta RV_{\mathrm{c}}$) with the depth of the 46 spectral lines selected (43 for $\tau^{1}$ Lup), one can directly measure, in principle, the atmospheric velocity gradient. In the case of $\alpha$ Lup, we find the following relation: 

\begin{equation}
\Delta RV_{\mathrm{c}} = -[0.24 \pm 1.23] D + [22.46 \pm 0.24] \:\mathrm{km\: s}^{-1} \, .
\end{equation}
The slope is consistent with a lack of velocity gradient within the atmosphere. For $\tau^{1}$~Lup, we find

\begin{equation} 
\Delta RV_{\mathrm{c}} = -[3.77 \pm 2.60] D + [8.50 \pm 0.40] \:\mathrm{km\: s}^{-1} \, ,
\end{equation}
which is consistent (at the 1.3$\sigma$ level) with a {\it negative} velocity gradient. Surprisingly, the amplitude of the radial velocity curves is larger near the photosphere (spectral lines of low depth) compared to the ones corresponding to the upper part of the atmosphere (spectral lines of large depth). 

Figure 6 shows the relation found for $\alpha$ Lup and $\tau^{1}$~Lup, together with previous results of Cepheids (Nardetto et al. 2007). The larger the 2K-amplitude, the larger the {\it observed} velocity gradient. This can also be seen in Fig. 7, where we show the slope $a_\mathrm{0}$ (from the $\Delta RV_{\mathrm{c}} = a_\mathrm{0} D + b_\mathrm{0}$ relations for all stars) as a function of the 2K-amplitude (i.e. $b_\mathrm{0}$). We find the following relation:

\begin{equation} 
a_\mathrm{0} = [0.25 \pm 0.01] b_\mathrm{0} -[5.79 \pm 3.05] 
\end{equation}

\begin{figure}[htbp]
\begin{center}
\resizebox{0.9\hsize}{!}{\includegraphics[clip=true]{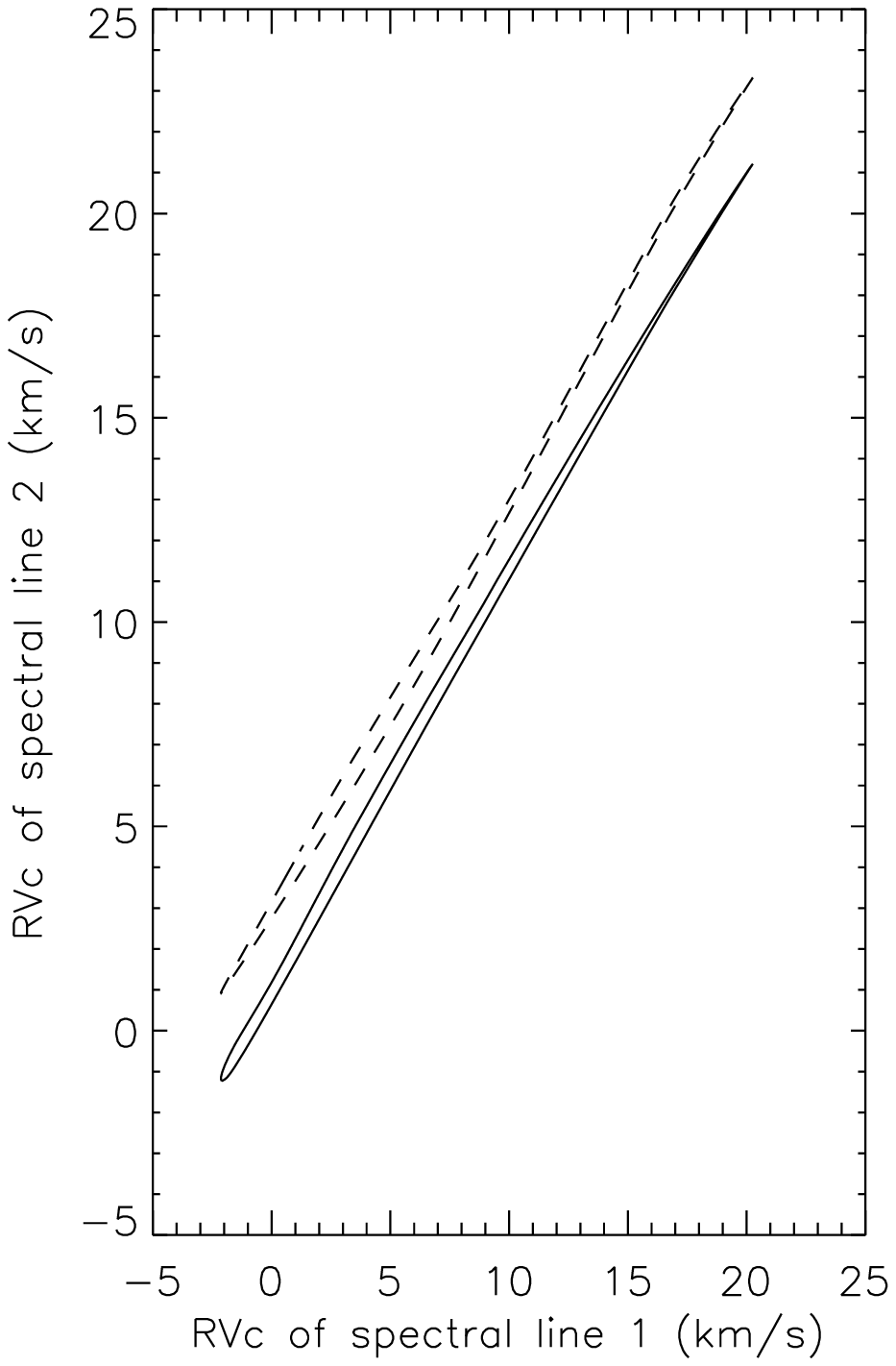}}
\end{center}
\caption{Van Hoof effect observed in the case of the $\beta$ Cephei star $\alpha$ Lup.} \label{Figuv}
\end{figure}
\begin{figure}[htbp]
\begin{center}
\resizebox{0.9\hsize}{!}{\includegraphics[clip=true]{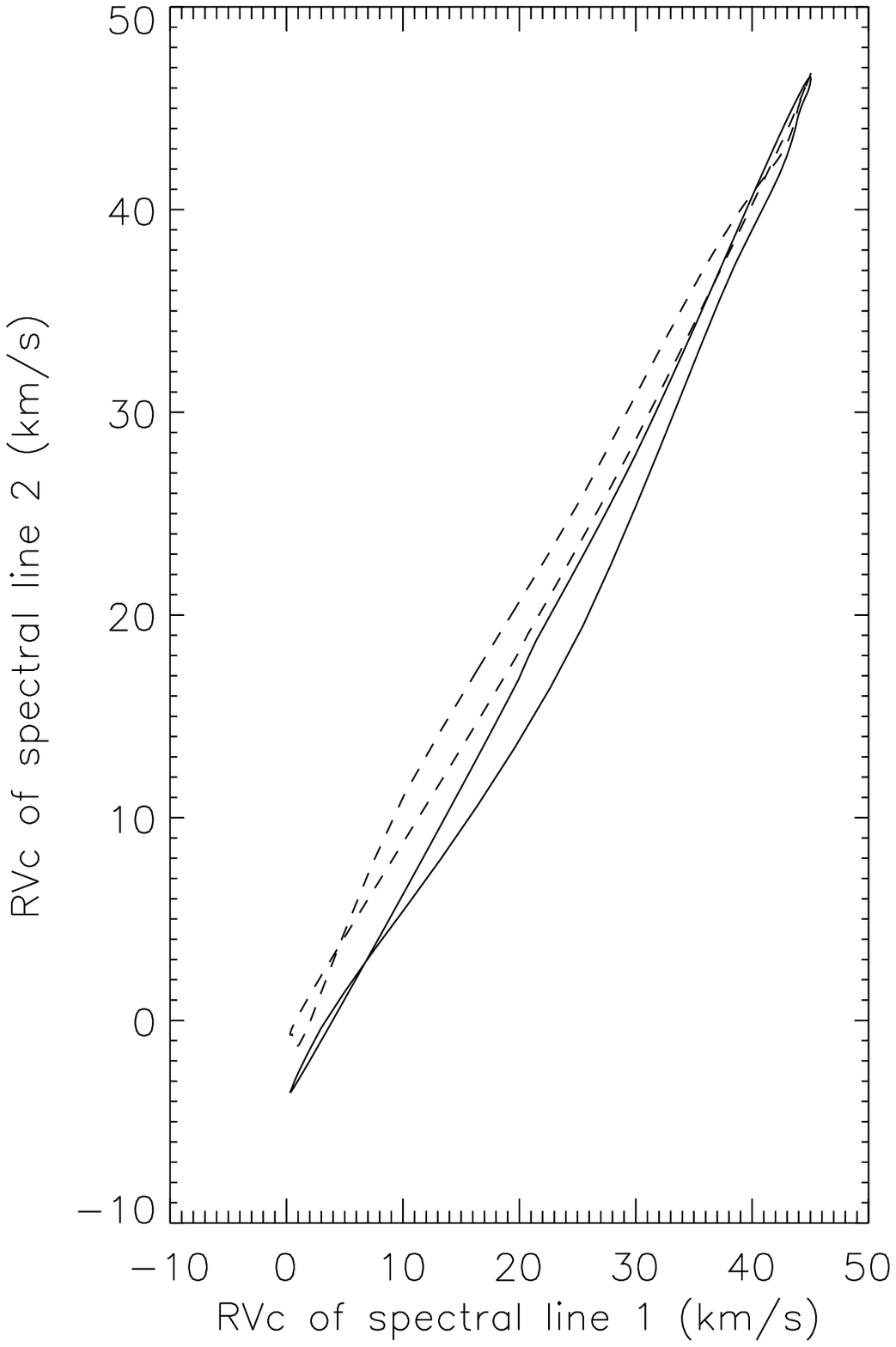}}
\end{center}
\caption{Van Hoof effect observed in the case of the Classical Cepheid RZ Vel.} \label{Figuv}
\end{figure}

This trend is interesting and brings intriguing new questions about the atmospheric velocity gradient of low-amplitude pulsating stars ($\Delta RV_{\mathrm{c}} < 20$~\kms) for which a {\it negative} velocity gradient is expected (as found for $\tau^{1}$~Lup). Is $\tau^{1}$Lup a peculiar star ? Shall we expect the same result for other kinds of low-amplitude pulsating stars ? Is it due to the presence of non-radial modes in the atmosphere of $\tau^{1}$ Lup ? More observations are needed to shed light on these questions. 

Another interesting approach is to study the Van Hoof effect, which is described in detail in Mathias \& Gillet (1993). Figure~8 shows the $RV_{\mathrm{c}}$ curves corresponding to spectral lines \ion{N}{II} 5679.5~\AA\  (solid line) and  \ion{O}{II} 4705.3~\AA\   (dashed line) as a function of the $RV_{\mathrm{c}}$ curves  of  \ion{C}{II} 6582.9~\AA\, in the case of $\alpha$~Lup. The two curves obtained show a loop shape due to the phase shifts between the three spectral lines considered (Van Hoof effect). We find that the two curves are parallel (because there is no velocity gradient observed from the radial velocity amplitudes) and shifted by a few~\kms (because of the difference in $\gamma$-velocities). For comparison, Figure~9 shows the same kind of plot but for the long-period Cepheid RZ Vel (HARPS observations presented in Nardetto et al. 2006a): $RV_{\mathrm{c}}$ curves corresponding to spectral lines \ion{N}{I} 5082.3~\AA\  (solid line) and  \ion{Si}{I} 6155.1~\AA\  (dashed line) are presented as a function of the $RV_{\mathrm{c}}$ curves  of  \ion{Fe}{I} 4896.4~\AA. In this case, the Van Hoof effect is significantly larger (larger loops) and the velocity gradient is clearly seen by the difference of slope between both curves. Again, we find a shift of a few~\kms due to the difference between the $\gamma$-velocities obtained for the two lines. 

These qualitative results show that such plots are extremely useful to synthesize most of the information concerning the dynamical structure of pulsating stars' atmosphere. In particular, in this case we find that a Van Hoof effect is possible even without an atmospheric velocity gradient.

\section{The semi-theoretical projection factor}

The basic approach used in Nardetto et al. (2004), where we provided a self-consistent projection factor of the Cepheid $\delta$~Cep using our non linear hydrodynamical model (Fokin 1990, 1991; Fokin et al. 2004) was not possible for $\alpha$~Lup. The reason is that the best model we obtain for $\alpha$ Lup does not reproduce the HARPS observations satisfactorily. The initial static model requires five input parameters: the luminosity ($L$), the effective temperature ($T_\mathrm{eff}$), the mass ($M$), and the chemical composition ($X$ and $Y$). The thus built static model is then perturbated, and its dynamic response is improved until the code converges toward a limit-cycle. After that, radiative transfer in the line is solved in the frame of this hydrodynamical model to provide line profiles (Fokin 1991). However, in this study the radiative transfer is not used. For $\alpha$~Lup, Zorec et al. (2009) gave an effective temperature for $\alpha$~Lup of $T_\mathrm{eff}=23100 \pm 1490$ K based on spectrophotometric data, while Heynderickx et al. (1994) found a mass of $M=10.4\Mo$. Considering these parameters and the HR diagram by Stankov \& Handler (2005, Fig. 5), we find a luminosity  of about $L=10000\Lo$. The metallicity is uncertain: Daszynska-Daszkiewicz \& Niemczura (2005) provide $[\frac{m}{H}]=Z=0.04\pm0.1$. We consider Y=0.294 and Z=0.006.  With the input parameters summarized in Table~4, the model converges toward the radial fundamental mode, with a pulsation period of $P=0.25$d (consistent at the 4\% level with observation), while the amplitude of the radial velocity curve is of about 9.5~\kms, which is more than twice lower than what we derived from observations ($\simeq 22$~\kms).  We tried to resolve this discrepancy by computing two other models with  $L=18000\Lo$ and  $L=25000\Lo$, the other parameters remained as they are in Table 4. Such luminosities increase the amplitude of the radial velocity curve to 12~\kms and 16~\kms, respectively. The discrepancy found between the observations and the model (in particular in terms of amplitude of the radial velocity curve) is not sufficient to derive a purely theoretical projection factor of $\alpha$~Lup. The fact that $\alpha$~Lup has an effective temperature and a luminosity about five times larger compared to for instance$\delta$ Cep, with an atmosphere five times less extended in percentage (0.4\% for $\alpha$ Lup and 2\% for $\delta$ Cep from our code) might play an important role in explaining this discrepancy. Originally, the code was developed for high-amplitude pulsating stars. The only $\beta$~Cepheid we successfully modelled was BW~Vul, which was presented in Fokin et al. (2004). Improving the model in order to reproduce $\alpha$ Lup observations remains an open problem to be resolved in future studies.

Nevertheless, even if this model of $\alpha$~Lup is imperfect, we can use the same approach as described in Guiglion et al. (2013) for $\delta$~Scuti stars and derive a semi-theoretical projection factor. There are, indeed, three separated concepts involved in the decomposition of the BW projection factor: $ p = p_\mathrm{0} f_\mathrm{grad} f_\mathrm{og}$. This expression has been  used for Cepheids (Nardetto et al. 2007; Nardetto et al. 2011). If we assume that this decomposition is correct for $\alpha$~Lup (which is a conservative approach), we can still provide a semi-theoretical projection factor for $\alpha$~Lup. In this case, our aim here is not to reproduce the spectral line profile or even the radial velocity curve, but to derive a few physical quantities associated with the photosphere of the star, such as $p_\mathrm{0}$ and $f_\mathrm{og}$.  The last quantity, $f_\mathrm{grad}$, is derived directly from our HARPS observations (as described below). Such an approach is enough to study the projection factor, at least at first order. 

First, the geometric projection factor, $p_\mathrm{0}$, is mainly related to the limb-darkening of the star. It corresponds to an integration of the pulsation velocity field (associated with the line-forming region) projected on the line of sight and weighted by the surface brightness of the star (including limb darkening within the spectral line). To derive $p_{\mathrm{o}}$, we consider two models: (1) the intensity distributions {\it in the continuum} provided by our hydrodynamical model (using parameters of Tab. 4) and (2) the corresponding static model of Claret et al. (2011). In both cases, we made the assumption that the limb-darkening variation within the line is negligible. The geometric projection factor derived from the hydrodynamical model varies from 1.435 (when considering the continuum next to the spectral line \ion{Si}{II} 3856.02 \AA) to 1.456 (continuum next to the \ion{C}{II} 6582.88 \AA\  spectral line).  We find a time independent limb darkening, with a mean value of $p_\mathrm{0}=1.44\pm0.01$ (average over the different wavelengths).  We use this value in the following. For $\delta$ Cep the time-variation of the limb darkening was also found negligible by Nardetto et al. (2006b).  Concerning the static model, the linear limb-darkening law of the continuum intensity profile of the star provided by Claret et al. (2000) and Claret \& Bloemen (2011) is $I(\cos(\theta))=1-u_{\mathrm V}+u_{\mathrm V}\cos(\theta)$, where $u_{\mathrm V}$ is the limb darkening of the star in V band and $\theta$ is the angle between the normal of the star and the line of sight. For Cepheids, $u_{\mathrm V} $ is close to $0.7$. For $\alpha$ Lup, using the physical parameters described in Tab. 4, we find $u_{\mathrm V}=0.37$, and using the relation linking $p_\mathrm{0}$ to  $u_{\mathrm V}$: $p_\mathrm{0}=\frac{3}{2}-\frac{u_{\mathrm V}}{6} $ (Getting 1934, Hadrava et al. 2009), we find $p=1.44$. We note that this value is consistent with Mathias et al. (1994) and with our hydrodynamical results, but is significantly larger compared to what we generally obtain for Cepheids (typically 1.36 to 1.41).

Second, the correction due to the velocity gradient within the atmosphere, $f_\mathrm{grad}$, is a quantity which can be directly derived from spectroscopic observations. Indeed, $f_{\text{grad}}$ depends on the spectral line considered: $f_{\text{grad}} = b_0 / (a_0D+b_0)$ (Nardetto et al. 2007, their Eq.~3). Using Eq.~6, we find that $f_{\text{grad}}$ is typically the same for all spectral lines ($f_{\text{grad}} = 1.00 \pm 0.01$), which is consistent with no correction of the projection factor due to the velocity gradient. 

The $f_{\text{o-g}}$ correction, which is the last component of the projection factor decomposition, cannot be measured from observations. To estimate the differential velocity between the \emph{optical} and \emph{gas} layers at the photosphere of the star, we need a hydrodynamic model (for a detailed definition of these two layers, we refer to Eq.~1 and~3 of Nardetto et al. 2004). For such determination, we use the nonlinear hydrodynamical model described above. We find $f_\mathrm{o-g}=0.99\pm0.01$ (numerical uncertainty derived from the model), whatever the model considered ($L=10000\Lo$, $L=18000\Lo$ or  $L=25000\Lo$). We note that $f_\mathrm{o-g}$ is mainly sensitive to the period (Nardetto et al. 2007). The impact of the amplitude of the radial velocity is negligible, which is confirmed by the fact that we find the same value for our three models.

We can now calculate the projection factor $p$, using the relation $p = p_\mathrm{0} f_\mathrm{grad} f_{\text{o-g}}$. We find $p  = 1.43 \pm 0.01$. 

\begin{table}
\begin{center}
\caption[]{Input parameters for the hydrodynamical modeling of the photosphere of $\alpha$~Lup used to derive $p_\mathrm{0}$ and $f_{\text{o-g}}$.}
\label{Tab_hydro}
\begin{tabular}{|c|c|}
\hline \hline \noalign{\smallskip}
   input parameters                                                         &    $\alpha$ Lup       \\                              
\hline
$L$  [$\Lo$] &  $10000$  \\ 
$M$  [$\Mo$] &  $12$  \\ 
$T_{\mathrm{eff}}$  [K] &  $23000$  \\ 
$X$                 &$ 0.700$  \\
$Z$                & $0.006$  \\
\hline \noalign{\smallskip}
\end{tabular}
\end{center}
\end{table}

\section{Conclusion}

Like any pulsating star, $\alpha$ Lup shows a cyclic variation of its spectral line asymmetry caused by its pulsation. Nevertheless, on average (i.e. over one pulsation cycle, and over all spectral lines in our sample) the spectral line asymmetry (or $\gamma$-asymmetry) is significantly  (at the 4$\sigma$ level) negative ($\simeq -7$\%). For Classical Cepheids, the $\gamma$-asymmetries are positive and range from about 0 to 5\%. This difference between $\beta$ Cephei stars and Classical Cepheids suggests that the different physical mechanisms at work in the atmospheric dynamics are still not all clarified. 

However, the $\beta$ Cephei stars have a dynamical structure of their atmosphere, which seems consistent with Classical Cepheids in the sense that we find a clear linear relation between the atmospheric velocity gradient and the amplitude of the radial velocity curve.  No velocity gradient has been measured in the atmosphere of $\alpha$ Lup, in contrast to Cepheids. In addition, $\tau^{1}$ Lup seems to have a {\it negative} velocity gradient. More observations of low amplitude pulsating stars are necessary to confirm this result. We also remark that the phase shift due to the Van~Hoof effect between two metallic lines forming at different levels in the atmosphere is also possible when the atmospheric velocity gradient is zero (as in the case of $\alpha$~Lup).

{Finally, by applying the usual decomposition of the projection factor to $\alpha$ Lup (which remains a conservative assumption), we find a semi-theoretical projection factor for $\alpha$ Lup of $p=1.43\pm0.01$. This value of the projection factor can be used with caution in a BW method of distance determination for this star. The Hipparcos parallax ($\pi = 7.02 \pm 0.17$ mas; Van Leeuwen 2007) and the radius estimate from Lesh \& Aizenman (1978), $R=10.8\Ro$ lead to an expected angular diameter of $0.7$ mas. Using our radial velocity curves (in particular, $2K=22.5$~\kms), the projection factor $p=1.43$, and the distance from Hipparcos, we find an absolute angular diameter variation of 4$\mu$as (or 0.6\%).  Unfortunately, there are no V, V-K photometric observations available for $\alpha$ Lup in the literature to confirm these results.

A comparison of different classes of pulsating stars in which radial modes dominate the pulsations seems an excellent way to gain an improved understanding of their respective dynamical structures, the k-terms, and the BW projection factors appropriate for each class of these variables. 

\begin{acknowledgements}
WG and GP gratefully acknowledge financial support for this work from the Chilean Center for Astrophysics FONDAP 15010003 and from the BASAL Centro de Astrofisica y Tecnologias Afines (CATA) PFB-06/2007. Support from the Ideas Plus programme of the Polish Ministry of Science and Higher Education and the TEAM subsidy of the Foundation for Polish Science (FNP) is also acknowledged.  This research has made use of the CDS Astronomical Databases SIMBAD. We thank the referee for his useful comments. \end{acknowledgements}



\begin{thebibliography}{}

\bibitem[2009]{Nardetto09} Balona, L. A. \& Feast, M. W. 1975, MNRAS, 172, 191
\bibitem[2009]{Nardetto09} Balona, L. A. \& Stobie, R. S. 1979, MNRAS, 187, 217

\bibitem[2000]{tbd00} Barbier-Brossat M., Figon P.  2000, A\&AS, 142, 217
\bibitem[2004]{tbd00} Tian, Bin, Men, Hui, Deng, Li-Cai, et al. 2003, Chin. J. Astron. Astrophys. Vol. 3, No. 2, 125
 \bibitem[2000]{tbd00} Breger, M. 1967, MNRAS, 136, 51
\bibitem[1996]{Butler96} Butler, R. P., Bell, R. A., Hindsley, R. B. 1996, ApJ, 461, 362
\bibitem[1996]{Butler96} Buscombe, W. \& Morris, P. M., 1960 MNRAS, 121, 263


\bibitem[1987]{Caldwell87}Caldwell, J. A. R. \& Coulson, I. M. 1987, AJ, 93, 1090
\bibitem[1938]{Camm38} Camm, G. L. 1938, MNRAS, 99, 71
\bibitem[1944]{Camm44} Camm, G. L. 1944, MNRAS, 104,163
\bibitem[1938]{Camm38} Campbell, W. W. 1928, Lick Obs., 16, 1
\bibitem[2000]{claret00} Claret, A. 2000, A\&A, 363, 1081
\bibitem[2000]{claret00} Claret, A. \& Bloemen, S. 2011, A\&A, 529, 75
\bibitem[2000]{claret00} Cuypers, J. 1987, A\&AS, 69, 445

\bibitem[2000]{claret00}  Daszynska-Daszkiewicz \& Niemczura 2005, A\&A, 433, 1031 
\bibitem[2004]{Kervella04} Davis, J., Jacob, A. P., Robertson, J. G. et al. 2009, AnGeo, 27, 2449
\bibitem[2004]{Kervella04} Dziembowski, W. 1977, A\&A, 27, 203 

\bibitem[2009]{Nardetto09} Evans, D. S. 1967, AUS, 30, 57

\bibitem[2004]{Kervella04} Kervella P., Nardetto N., Bersier D., et al. 2004, \aap, 416, 941

\bibitem[1995]{Fernie95} Fernie, J.D., Beattie, B., Evans, N.R., and Seager, S. 1995, IBVS No. 4148
\bibitem[1990]{f90} Fokin A.B. 1990, \apss, 164, 95
\bibitem[1991]{f91} Fokin A.B. 1991, \mnras, 250, 258
\bibitem[2004]{tbd00}Fokin, A., Mathias, Ph., Chapellier, E., et al.  2004, A\&A, 426, 687
\bibitem[2007]{Fouque07} Fouqu\'e, P., Arriagada, P., Storm, J. et al. 2007, A\&A, 476, 73

\bibitem[1998]{gieren98} Getting 1934, MNRAS, 95, 139
\bibitem[1998]{gieren98} Gieren, W. P., Fouqu\'e, P., \& G\'omez, M. 1998, ApJ, 496,17
\bibitem[2005]{gieren05} Gieren, W. P., Storm, J., Barnes, T. G., et al. 2005a, ApJ, 627, 224
\bibitem[2005]{gieren05} Gieren, W., Pietrzynski, G., Bresolin, F., et al. 2005b, Msngr, 121, 23
\bibitem[2005]{gieren05} Guiglion, G., Nardetto, N., Mathias, P., et al. 2013, A\&A, 550, 10

\bibitem[2005]{gieren05} 	Hadrava, P., Slechta, M., Skoda, P. 2009, A\&A, 507, 397
\bibitem[2005]{gieren05} 	Hatzes, Artie P. 1996, PASP, 108, 839	
\bibitem[2005]{gieren05}Heiter, U, Barklem, P., Fossati, L. et al. 2008 J. Phys.: Conf. Ser. 130, 012011
\bibitem[2004]{tbd00} Heynderickx, D. 1992, A\&AS, 96, 207
\bibitem[2004]{tbd00} Heynderickx, D.,  Waelkens, C., Smeyers, P., D. 1994, A\&AS, 105, 447


\bibitem[2009]{Nardetto09} Jakate, S. M. 1980,  A\&A, 84, 374
\bibitem[1996]{Butler96} Joy, A. H. 1939, ApJ, 89, 356

\bibitem[2004]{tbd00} Lampens, P., Goossens, M. 1982, A\&A 115, 413 
\bibitem[2009]{Nardetto09} Lesh \& Aizenman 1974, A\&A, 34, 203
\bibitem[2009]{Nardetto09} Leung, K. C. 1967, ApJ, 150, 233


\bibitem[2009]{Nardetto09} McNamara \& Mathews 1967, mamt. book. . 127
\bibitem[2009]{Nardetto09} Mathias, P., Gillet, D., Crowe, R. 1991, A\&A, 252, 245
\bibitem[2009]{Nardetto09} Mathias, P. \& Gillet, D. 1993, A\&A, 278, 511
\bibitem[2004]{tbd00} Mathias, P., Aerts, C., De Pauw, M., et al. 1994, A\&A, 283, 813 
\bibitem[2004]{tbd00} Mathias, P., Gillet, D., Fokin, A. B. et al. 2006, A\&A, 457, 575 
\bibitem[2005]{Merand05} M\'erand A., Kervella P., Coud\'e du Foresto V. et al. 2005 \aap, 438, 9 
\bibitem[2004]{tbd00} Milone, L. 1962, Asso. Argent. de Astron. Bul. 4, 42
\bibitem[1987]{Moffett87} Moffett, T. J., Barnes, T. G. III 1987, PASP, 99, 1206M

\bibitem[2004]{Nardetto04} Nardetto, N., Fokin, A., Mourard, D.,  et al. 2004,  A\&AP, 428, 131
\bibitem[2004]{Nardetto04} Nardetto, N., Mourard, D., Kervella, P. et al. 2006a,  A\&A, 453, 309
\bibitem[2004]{Nardetto04} Nardetto, N., Fokin, A., Mourard, D. et al. 2006b,  A\&A, 454, 327
\bibitem[2007]{Nardetto07} Nardetto, N., Mourard, D., Mathias, P. et al. 2007, A\&A, 471, 661
\bibitem[1996]{Butler96} Nardetto, N., Stoekl, A., Bersier, D. et al.  2008, A\&A, 489, 1255
\bibitem[2009]{Nardetto09} Nardetto, N., Gieren, W., Kervella P. et al. 2009, A\&A, 502, 951
\bibitem[2009]{Nardetto09} Nardetto, N., Fokin., A., Fouqu\'e, P. et al. 2011, A\&A, 534, 16 

\bibitem[2004]{tbd00} Pagel, B. E. J. 1956, MNRAS 116, 10
\bibitem[1945]{Parenago45} Parenago, P. P. 1945, PA, 53, 441
\bibitem[1945]{Parenago45} Piskunov N.E., Kupka F., Ryabchikova T.A. et al. 1995, A\&AS 112, 525

\bibitem[1994]{Pont94} Pont, F., Mayor, M., \& Burki, G. 1994, A\&A, 285, 415

\bibitem[2004]{tbd00} Reed, B. Cameron 2003, AJ, 125, 2531
\bibitem[2004]{tbd00} Rodgers, A. W., Bell, R. A. 1962, Observatory, 82, 26

\bibitem[1995]{Sabbey95} Sabbey, C. N., Sasselov, D. D., Fieldus, M. S, et al. 1995, \apj, 446, 250
\bibitem[1995]{Sabbey95} Stamford, P. A. \& Watson, R. D. 1981, Ap\&SS, 77, 131
\bibitem[2009]{Nardetto09} Stankov, A, Handler, G 2005, ApJS, 158, 193
\bibitem[2009]{Nardetto09} Sterken, C. \& Jerzykiewicz M. 1993, SSRv, 62, 95
\bibitem[1956]{Stibbs56} Stibbs, D. W. N. 1956, MNRAS, 116, 453
\bibitem[2004]{tbd00}Storm J., Gieren W., Fouqu\'e P.  2011a, A\&A, 534, 94
\bibitem[2004]{tbd00}Storm J., Gieren W., Fouqu\'e P.  2011b, A\&A, 534, 95

\bibitem[1956]{Stibbs56} Thackeray, A. D. 1966, MnRAS, 70, 33

\bibitem[2009]{Nardetto09} Waelkens, C. 1981, A\&A, 97,274

\bibitem[2004]{tbd00} Van Hoof, A. 1964, Z. f. Ap. 224, 953
\bibitem[2004]{tbd00} Van Hoof, A. 1965, Kl. Veršf. d. Remais-Sternwarte Bamberg 4, n¡ 40, 149
\bibitem[2004]{tbd00} Van Leeuwen F.  2007, A\&A, 474, 653

\bibitem[1974]{Wielen74} Wielen, R. 1974, A\&AS, 15, 1
\bibitem[1953]{Wilson91} Wilson, R. E. 1953, GCRV, C, 0
\bibitem[1991]{Wilson91} Wilson, T. D., Barnes, T. G., Hawley, S. L. \& Jefferys, W. H. 1991, ApJ, 378, 708
\bibitem[1991]{Wilson91} Wright, W. H. 1909, Lick Obs., 5, 176
\bibitem[1991]{Wilson91} Wright, W. H. 1911, Lick Obs., 9, 71

\bibitem[2009]{Zima08} Zima, W. 2008, Communications in Asteroseismology 155
\bibitem[2009]{Zorec09}  Zorec, J., Cidale, L.,Arias, M. L. 2009, A\&A, 501, 297
	
\end{thebibliography}
\end{document}